\documentclass{article} % For LaTeX2e
\usepackage[submission]{colm2025_conference}

\colmsubmissionfalse
\colmpreprintfalse
\colmfinaltrue

\usepackage{microtype}
\usepackage{hyperref}
\usepackage{url}
\usepackage{booktabs}
\usepackage{tabularx}
\usepackage{array}
\usepackage{placeins}
\usepackage[T1]{fontenc}

\usepackage{lineno}

\usepackage{minted}
\usepackage{amsmath,amssymb}
\usepackage{tikz}
\usetikzlibrary{positioning,fit,shapes.geometric,arrows.meta}
\usepackage{listings}
\usepackage{xcolor}
\usepackage{algorithm}
\usepackage{algpseudocode}
\usepackage{titlesec}
\usepackage{graphicx}

\usepackage{verbatim}
\usepackage{enumitem}

\definecolor{codebg}{RGB}{248,248,248}
\definecolor{codeframe}{RGB}{220,220,220}
\definecolor{dkgreen}{rgb}{0,0.6,0}
\definecolor{gray}{rgb}{0.5,0.5,0.5}
\definecolor{mauve}{rgb}{0.58,0,0.82}

\setlist{nosep}                % global: zero \itemsep, \topsep, \partopsep

\definecolor{darkblue}{rgb}{0, 0, 0.5}
\hypersetup{colorlinks=true, citecolor=darkblue, linkcolor=darkblue, urlcolor=darkblue}

\title{NegotiationGym: Self-Optimizing Agents in a Multi-Agent Social Simulation Environment}

\author{Shashank Mangla*, Chris Hokamp*, Jack Boylan, Demian Gholipour Ghalandari, \\
\textbf{Yuuv Jauhari, Lauren Cassidy, Oisin Duffy} \\
Quantexa \\
\texttt{<firstname><lastname>@quantexa.com} \\}
\begin{document}

\ifcolmsubmission
\linenumbers
\fi

% \colmfinalcopy

\maketitle

\def\thefootnote{*}\footnotetext{equal contribution}\def\thefootnote{\arabic{footnote}}

\begin{abstract}
We design and implement NegotiationGym -- an API and user interface for configuring and running multi-agent social simulations focused upon negotiation and cooperation. The NegotiationGym codebase offers a user-friendly, configuration-driven API that enables easy design and customization of simulation scenarios. Agent-level utility functions encode optimization criteria for each agent, and agents can self-optimize by conducting multiple interaction rounds with other agents, observing outcomes, and modifying their strategies for future rounds. 
\end{abstract}
% \yuuv{above, a bit more explanation what the agents are optimizing towards? better prompts/utility functions}
% \chris{experiments or case studies?} 
% Agents can self-optimize using their goals or utility functions which may be at odds, especially in zero-sum settings. By optimizing the prompts of single agents, we test the impact of individual outcomes (e.g. buyer surplus), broader system effects (e.g. seller profit, deal rate, conversation length), and robustness across LLM stochasticity (variance of outcomes for the system). 

\section{Introduction}

Rational actors weigh the expected cost of actions when making decisions under uncertainty \citep{ClarkeCornishrationalchoicetheorybook}. In social interactions between humans, complex dynamics emerge through observation of and reaction to the behavior of others. Especially in interactions where individuals have clear objectives, the strategy employed by each individual can drastically affect the final outcome. Through experience, humans can develop advanced strategies that improve performance, and more experience often correlates with better strategic performance. 

AI agents can be used to model human behavior, and modern general-purpose LLMs enable simulations of complex social scenarios where multiple agents interact. Although various works have modeled multi-agent scenarios using Agentic AI \citep{li2023camelcommunicativeagentsmind, xiao2025tradingagentsmultiagentsllmfinancial, du2023improvingfactualityreasoninglanguage}, there remains a lack of flexible frameworks and abstractions that allow researchers to easily design and run simulations.
This work introduces \texttt{NegotiationGym}, an open-source toolkit that makes it easy to configure and run multi-agent simulations, measure outcomes, and optimize agents. 

Simulations in \texttt{NegotiationGym} host a flexible number of LLM agents, taking roles such as \textit{seller} and \textit{buyer}. Agents can be assigned fixed strategies for a given episode and are evaluated using dedicated fitness or utility functions after each simulation run. An experiment harness allows users to run particular simulation scenarios multiple times, with the ability to optimize agents to improve their individual performance.

% Current Agentic AI systems enable configurable simulations of interactions between agentic entities with individual goals. In real-world scenarios, interacting entities have access to both private information and shared information about the environment.
% in effect, every entity keeps an internal model of every other entity
% use all available information to optimize their performance relative to a set of utility functions.

% For example, in a buyer-seller simulation, a buyer agent may try to acquire an item for the lowest possible price, subject to a private maximum budget and a learned negotiation strategy, while a seller agent tries to negotiate for the highest price, subject to a private minimum and a sales strategy.\chris{reference figure}.

% \begin{itemize}
%     \item A flexible number of LLM agents that take on different roles such as seller and buyer
%     \item Agent-level fixed \textbf{strategies} for each episode
%     \item A \textbf{fitness/utility function} for each agent which measures performance after a simulation run
%     \item An experiment harness that runs a particular simulation scenario $k$ times, with options to optimize the utility of individual agent's across runs by revising their prompts
%     % \item Dynamic input from the environment 
% \end{itemize}
% in the case of LLMs, the execution of the strategy is affected by the agent's prompt, and the trained model's parameters.
In our example scenario, we find non-trivial improvement (e.g., sizable utility/surplus shifts and fewer no-deals) when agents learn from feedback across simulation runs. We chose to focus solely on utility-based prompt feedback because utility-based feedback offers clear, quantifiable grounding for agent optimization. Our experiments focus on negotiation scenarios and study the effect of optimizing buyer and seller agents across runs (Section \ref{sec:experiments}).

The key contributions of this work are: 1) A simple, extensible framework for multi-agent simulations; 2) A flexible interface for configuration of agent goals, private information, and stop conditions; and 3) A case study with analysis for a sale negotiation scenario.

The paper is organized as follows: Section \ref{sec:related-work} discusses essential related work, Section \ref{sec:negotiation-gym} presents \texttt{NegotationGym}, Section \ref{sec:experiments} presents an example experiment conducted using the framework, and Sections \ref{sec:limitations} and \ref{sec:conclusion} discuss limitations and conclusions.

\section{Related Work}
\label{sec:related-work}

% \shashank{I have commented out the first paragraph here and removed paragraph titles to reduce this section}

% \paragraph{LLMs for task execution}
% Prompt engineering has pushed LLMs beyond simple chat-based use cases to solving complex tasks that require careful thought, planning, and execution. 
% \cite{wei2023chainofthoughtpromptingelicitsreasoning} 
% shows that providing step-by-step reasoning examples prompts LLMs to generate reasoning traces and improves performance.
% \cite{yao2023reactsynergizingreasoningacting} extend this idea to interleave the generation of reasoning traces and actions (e.g., web search). Advances in prompt engineering form the foundation for our simulation environment, which leverages optimized prompts to enable complex, interactive agent behaviors.

% \paragraph{Multi-agent systems}
Multi-agent LLM settings can simulate collaborative or competitive social scenarios. 
\cite{park2023generativeagentsinteractivesimulacra} deploy LLM agents in a sandbox town environment where agents can access memory, plan, and interact, leading to emerging social behaviors (e.g., information diffusion, relationship formation, and coordination).
\cite{sreedhar2024simulatinghumanstrategicbehavior} show that multi-agent setups simulate human behavior in a game more accurately than single-agent setups.
\cite{xiao2025tradingagentsmultiagentsllmfinancial} propose a multi-agent trading framework, where agents assuming specialized roles collaborate and execute trade decisions. Systems can also have competitive agents: \cite{du2023improvingfactualityreasoninglanguage} use debating agents to converge on the best solutions.
Our framework can simulate similar settings and quantify how prompt optimization can measurably shift outcomes.

% \paragraph{Buyer-seller simulations}
Several recent works simulate buyer-seller scenarios. \cite{zhu2025automatedriskygamemodeling} show that agent negotiations become economically imbalanced when LLMs have differing capabilities that lead the weaker agent to concede and incur an economic loss. They also find that delegating negotiation to LLMs introduces risks such as budget constraint violation and excessive overpayment. \cite{oh2025llmagentsbargainingutilitybased} find that the LLM agent-based buyer tactics do not align with human norms, resulting in suboptimal negotiation. They introduce a feedback mechanism that allows agents to estimate their utility and adjust actions mid-negotiation. Our study extends such analyses and facilitates further research on the impact of agent optimization on economic outcomes and negotiation dynamics.
% \oisin{Maybe mention this paper: https://arxiv.org/pdf/2402.08189 shows how agentic frameworks are better than LLMs at replicating human behaviour in strategy-based sims}

% \paragraph{Self-optimizing agents}
Several works show that LLMs can improve without gradient updates by using self-improving feedback loops and learning from past experiences on the fly. Our framework draws on these concepts of self-optimization to iteratively enhance agent strategies. \cite{shinn2023reflexion} introduces a policy optimization technique that uses external reward signals, internal evaluation, and verbal self-reflection feedback stored in memory to improve the performance of an LLM agent in subsequent tests. However, \cite{huang2023large} evaluated several self-correcting techniques and found that intrinsic self-correcting techniques that do not use external oracle signals actually decrease reasoning performance. \cite{fu2023improving} used an LLM critic to provide feedback to agents in a buyer-seller simulation. Feedback is generated after negotiations and added as an update to the agent’s prompt for subsequent trials. We adopt this feedback loop along with utility-based prompt optimization in a case study for our framework and measure its effect.

% \chris{essential related work: https://arxiv.org/pdf/2410.13915 https://github.com/sandbox-social/mastodon-sim}

% \chris{see linked papers in the workshop site, make sure relevant papers are included https://sites.google.com/view/social-sims-with-llms}

\section{NegotiationGym}
\label{sec:negotiation-gym}

The framework is implemented using AutoGen\footnote{\url{https://ag2.ai/}} \citep{wu2023autogenenablingnextgenllm}. Unlike prior uses of AutoGen focused on general multi-agent coordination, our framework extends it with utility-aware agents, scenario-specific optimization hooks, and a configurable interface for iterative, outcome-driven negotiation simulations. \texttt{NegotiationGym} includes both a CLI and a GUI for configuring, running, and analyzing simulations\footnote{All code, prompts, and experiment configurations are available here \url{https://github.com/chrishokamp/multi-agent-social-simulation}}. 

% Simulations can be orchestrated directly using the APIs and provided scripts, or by using the user-interface, which provides an end-to-end workflow for configuration, running, and analyzing results

When simulations are configured in the GUI, an orchestrator streams jobs from a
MongoDB-backed queue, enabling multiple users to use the same backend deployment. Each job is controlled by \texttt{SelectorGCSimulation}, which controls the simulation components:

\begin{enumerate}
\item A \textbf{SelectorGroupChat} (AutoGen) that maintains the shared
history~\(H\) and chooses which agent acts next.
\item \textbf{Termination conditions}, which determine when an episode has finished. These may be triggered by agents themselves, or by environment constraints such as time or maximum number of messages.
\item A lightweight \textbf{environment} object
\(\mathbf{E}=\{\texttt{runs}:[\dots]\}\) that accumulates each finished run in
chronological order.  The object is passed to every agent after each episode,
enabling both utility computation and prompt revision.
\end{enumerate}

\paragraph{Configuration}
\label{subsec:configuring-agents}
Agents are declared in JSON and mapped to Python classes at runtime. Figure \ref{fig:config-example} in the appendix shows an example simulation configuration file. 

At runtime, every agent dictionary is converted to a \texttt{UtilityAgent}.  The base class instantiates an agent with two hooks:

\begin{itemize}
\item \textbf{\texttt{compute\_utility(\,E\,)}} – returns a scalar utility with respect to the agent's strategy and goals.  The default implementation returns 0.
\item \textbf{\texttt{learn\_from\_feedback(\,E\,)}} – may rewrite its own
\texttt{prompt} leveraging observations from the traceback of previous runs
(see \S\ref{subsec:optimization}).
\end{itemize}

Because the core framework never inspects an agent’s private data, users can
invent additional subclasses simply by overriding these hooks.

\paragraph{Running Simulations and Observing Outcomes}

\texttt{NegotiationGym} can be run from the command-line or GUI. A simulation runs for a configurable number of episodes. After each episode, configuration determines agents' utility functions and optimization requirements for future episodes. At the end of a simulation, a detailed report of outcomes is saved, and plots and other data are accessible via an interactive GUI.
% \yuuv{Maybe insert a screencap of the render?}
% Chris: template for inserting code (alternative to frame used earlier)
% \begin{minted}[fontsize=\small,linenos]{python}
%     # code here
% \end{minted}

\paragraph{Optimization}
\label{subsec:optimization}

Agents can perform a self-optimization procedure by invoking their overridable \texttt{optimize} function. The default implementation (a) assembles the
last ten episodes into a \emph{reflection prompt} (Appendix \ref{fig:reflection-prompt}), (b) asks the back-end LLM
to rewrite the current \texttt{system\_prompt} so as to increase the measured
utility given the agent’s private \texttt{strategy}, and (c) replaces the old
prompt in place.  Because the simulation passes the augmented environment~\(E\)
into the \emph{next} episode, the updated prompt immediately affects behavior.

Agents are therefore free to pursue heterogeneous learning rules—gradient-free
prompt search, bandit algorithms, or offline fine-tuning, as long as those rules
are encapsulated in \textbf{\texttt{learn\_from\_feedback(\,E\,)}}. This separation keeps the core simulation loop agnostic to optimization details while enabling
research on autonomous strategy improvement.

% \subsection{Visualizing Simulation Results}
% \subsection{Evaluation}
% \label{subsec:evaluation}

% \texttt{NegotiationGym} includes rich analysis and visualization tooling

% \chris{UI outcomes screenshot in appendix}

% \chris{add data science (make run-simulation) API tooling screenshot or reference}

\section{Case Study: Buyer–Seller Negotiation Coaching}
\label{sec:experiments}

% We study a series of controlled simulations, focusing upon extracting insights into optimization behavior, and the capabilities and limitations of different LLM model backends. We create both adversarial and co-operative \lauren{conflicting vs mutual interests} scenarios, and confirm that agents demonstrate the expected behavior in these environments. We then allow individual agents to self-optimize by observing the outcomes of multiple simulation rounds, and study whether this observation leads to improved outcomes for the optimizing agent in later rounds. In Section \ref{sec:analysis} we analyze the reasons why agents may succeed or fail.

\begin{figure}[t]                
  \centering
  \includegraphics[width=\linewidth]
                  {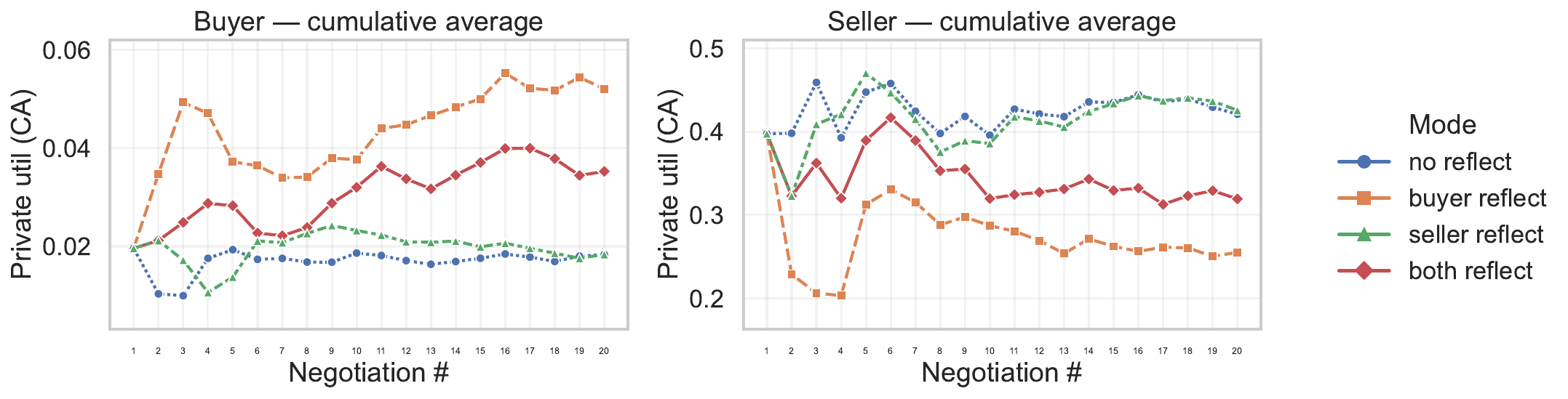}
  \caption{Cumulative average utility curves for each optimization mode on a 20-turn setting.}
  \label{fig:priv-utils}
\end{figure}

\begin{figure}[t]                      
  \centering
  \includegraphics[width=0.8\linewidth] 
                   {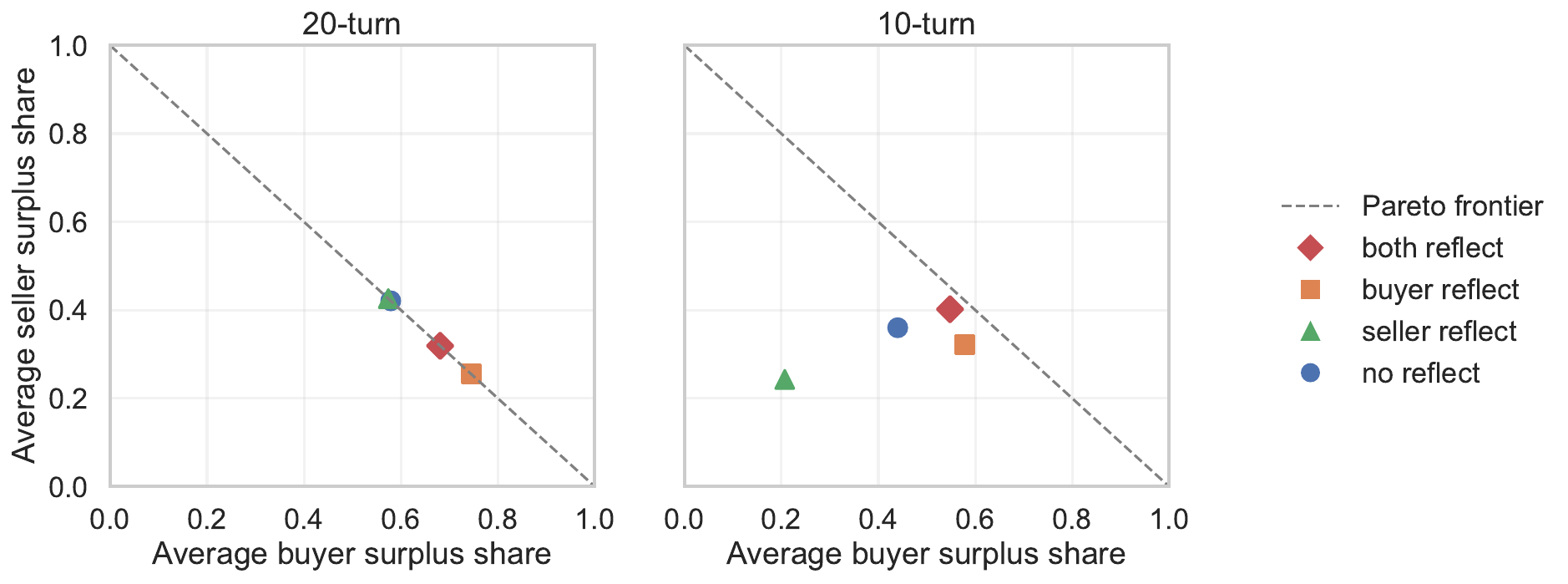}
  \caption{Average surplus shares over 20 negotiations for each optimization mode
           on a 20-turn setting (left) and a 10-turn setting (right).
           Points lie on or below the Pareto frontier. All prompts and configuration for this experiment are available in the open-source repository referenced above.}
  \label{fig:pareto-compare}
\end{figure}

We use \texttt{NegotiationGym} to set up a buyer and a seller agent negotiating the sale of a laptop. Each agent has a private utility function. A \textit{negotiation coach} agent is available that can analyze the transcript of a previous negotiation and an agent's utility and privately suggest negotiation strategies to either agent for future negotiations. We evaluate four modes, no-reflect (neither coached), buyer-reflect (only buyer is coached), seller-reflect (only seller is coached), and both-reflect (both coached). We used the GPT-4o model provided by OpenAI. The \textit{negotiation coach} agent prompt is shown in Appendix \ref{fig:negotiation-coach-prompt}.

Each negotiation samples a seller asking price, \(\text{ask}\sim U(900,1400)\,\mathrm{USD}\); draws the seller’s private floor as, \(\text{floor}=\text{ask}-U(100,300)\); and sets the buyer’s private budget as \(\text{budget}\sim U\!\bigl[\text{floor}+50,\;\text{ask}-50\bigr].\) The private utilities of the buyer and seller are defined as,
\[
u_{\text{buyer}}^{\text{priv}}
      = \frac{\text{budget}\;-\;p}{\text{budget}},
\qquad
u_{\text{seller}}^{\text{priv}}
      = \frac{p\;-\;\text{floor}}{\text{ask}\;-\;\text{floor}},
\]

where $p$ is the final price agreed upon.

We run the simulation for 20 negotiations, with a maximum of 20 turns each. If an agent is being coached, the negotiation coach analyses the transcript and utility at the end of each negotiation and appends a negotiation strategy to the agent's system prompt, that may then be utilized in the next negotiation.

Figure \ref{fig:priv-utils} shows the cumulative average of private utilities over the 20 negotiations for all four modes. We observe the highest cumulative utility for the buyer and lowest for the seller in the buyer-reflect mode. The opposite is the case in seller-reflect mode, although the seller utility is only marginally improved over the no-reflect mode. The both-reflect mode balances the utility of both agents. 

We also look at how the average surplus, the money that lies between the seller's private floor and the seller's public asking price, gets divided between the agents after negotiations. The buyer and seller surplus share for a single negotiation are defined as,
\[
\text{buyer\_ss}  = \frac{\text{ask}-p}{\text{ask}-\text{floor}}, 
\qquad
\text{seller\_ss} = \frac{p-\text{floor}}{\text{ask}-\text{floor}}.
\]

In a perfect zero-sum game, $\text{buyer\_ss} + \text{seller\_ss} = 1$. However, if a deal isn't reached in the maximum number of turns, we label it no-deal and assign both agents zero utility and surplus. Figure \ref{fig:pareto-compare} compares average surplus share over 20 negotiations for two settings, (a) maximum 20 turns; and (b) maximum 10 turns. With 20-turns, all negotiations reach a deal and all surplus is utilized. With 10-turns, all modes have no-deal cases, leaving unclaimed surplus. While the seller-reflect mode shifts surplus towards the seller, it causes many no-deals, leaving the most value unclaimed. The both-reflect mode, where both agents are optimized, has the fewest no-deals, indicating that agents learn to close deals fast and minimize net surplus loss as an emergent by-product, given they are unaware of the maximum turn-limit.

We observe in Figure \ref{fig:priv-utils} that the buyer gains more from feedback-driven optimization than the seller. One explanation is that the buyer's negotiation position is inherently more flexible: they can explore a wide range of counteroffers and concession strategies without risking a failed deal, whereas the seller, anchored by their floor price, has less room to maneuver. Additionally, buyers may gain more from learning timing and anchoring tactics (e.g., delaying concessions, reframing value), which translate directly into improved utility under our experimental setup. This asymmetry highlights the importance of role-specific optimization dynamics, and motivates the need for role-aware learning strategies in future work.

% \section{Analysis}
% \label{sec:analysis}
% \shashank{Should we delete this section or move the last paragraphs of Section 4 in here?}

\section{Limitations}
\label{sec:limitations}
Simulation outcomes are stochastic, so results can vary between runs and may need to be averaged over many runs to increase the reliability of conclusions. We use a simple price-based utility in our case study, but real-world utilities may be multifaceted, encompassing many other factors. Current simulations lack real-world grounding, and future extensions such as tool-use would enable agents to query external sources to align tactics to factual data. Agent behavior is also fully dependent upon the model(s) being used to run the simulation --- we did not investigate other models beyond GPT-4o, but we expect significant differences between models, with performance dependent both upon overall model quality, and upon models' ability to follow behavioral instructions in agents' prompts, which may be at odds with some models' default personalities. 

\section{Conclusion}
\label{sec:conclusion}

We have presented \texttt{NegotiationGym}, a configurable multi-agent simulation environment that allows exploration of complex social scenarios, and to optimize the utility of individual agents, which we have explored in a case study. The framework is simple to install and to extend to custom scenarios.

% \section{Ethics Statement}
% \label{sec:ethicsstatement}

% % \section{Author Contributions}
% % \label{sec:authorcontributions}

\section{Acknowledgments}
\label{sec:acknowledgments}
We acknowledge the significant contribution of the TCD team (Trinity College Dublin Software Engineering 2025 Group 22), who, under the supervision of our NLP team at Quantexa, kick-started the development of this project.
We are also grateful to the anonymous reviewers for their helpful and constructive feedback.

% % \chris{ping the team before the final version}

% \section*{Acknowledgments}
% Use unnumbered first level headings for the acknowledgments. All

% \section*{Ethics Statement}
% Authors can add an optional ethics statement to the paper. 
% For papers that touch on ethical issues, this section will be evaluated as part of the review process. The ethics statement should come at the end of the paper. It does not count toward the page limit, but should not be more than 1 page. 

\bibliography{colm2025_conference}

\begin{thebibliography}{15}
\providecommand{\natexlab}[1]{#1}
\providecommand{\url}[1]{\texttt{#1}}
\expandafter\ifx\csname urlstyle\endcsname\relax
  \providecommand{\doi}[1]{doi: #1}\else
  \providecommand{\doi}{doi: \begingroup \urlstyle{rm}\Url}\fi

\bibitem[AL et~al.(2024)AL, Ahn, Becker, Carroll, Christie, Cortes, Demirci, Du, Li, Luo, Wang, Willows, Yang, and Yang]{altera2024projectsidmanyagentsimulations}
Altera. AL, Andrew Ahn, Nic Becker, Stephanie Carroll, Nico Christie, Manuel Cortes, Arda Demirci, Melissa Du, Frankie Li, Shuying Luo, Peter~Y. Wang, Mathew Willows, Feitong Yang, and Guangyu~Robert Yang.
\newblock Project sid: Many-agent simulations toward {AI} civilization, 2024.
\newblock URL \url{https://arxiv.org/abs/2411.00114}.

\bibitem[Cornish \& Clarke(2017)Cornish and Clarke]{ClarkeCornishrationalchoicetheorybook}
D.B. Cornish and R.V. Clarke.
\newblock \emph{The Reasoning Criminal: Rational Choice Perspectives on Offending}.
\newblock 09 2017.
\newblock ISBN 9781315134482.
\newblock \doi{10.4324/9781315134482}.

\bibitem[Du et~al.(2023)Du, Li, Torralba, Tenenbaum, and Mordatch]{du2023improvingfactualityreasoninglanguage}
Yilun Du, Shuang Li, Antonio Torralba, Joshua~B. Tenenbaum, and Igor Mordatch.
\newblock Improving factuality and reasoning in language models through multiagent debate, 2023.
\newblock URL \url{https://arxiv.org/abs/2305.14325}.

\bibitem[Fu et~al.(2023)Fu, Peng, Khot, and Lapata]{fu2023improving}
Yao Fu, Hao Peng, Tushar Khot, and Mirella Lapata.
\newblock Improving language model negotiation with self-play and in-context learning from {AI} feedback.
\newblock \emph{arXiv preprint arXiv:2305.10142}, 2023.

\bibitem[Huang et~al.(2023)Huang, Chen, Mishra, Zheng, Yu, Song, and Zhou]{huang2023large}
Jie Huang, Xinyun Chen, Swaroop Mishra, Huaixiu~Steven Zheng, Adams~Wei Yu, Xinying Song, and Denny Zhou.
\newblock Large language models cannot self-correct reasoning yet.
\newblock \emph{arXiv preprint arXiv:2310.01798}, 2023.

\bibitem[Li et~al.(2023)Li, Hammoud, Itani, Khizbullin, and Ghanem]{li2023camelcommunicativeagentsmind}
Guohao Li, Hasan Abed Al~Kader Hammoud, Hani Itani, Dmitrii Khizbullin, and Bernard Ghanem.
\newblock Camel: Communicative agents for "mind" exploration of large language model society, 2023.
\newblock URL \url{https://arxiv.org/abs/2303.17760}.

\bibitem[Oh et~al.(2025)Oh, Aghazada, Yun, and Kim]{oh2025llmagentsbargainingutilitybased}
Jihwan Oh, Murad Aghazada, Se-Young Yun, and Taehyeon Kim.
\newblock {LLM} agents for bargaining with utility-based feedback, 2025.
\newblock URL \url{https://arxiv.org/abs/2505.22998}.

\bibitem[Park et~al.(2023)Park, O'Brien, Cai, Morris, Liang, and Bernstein]{park2023generativeagentsinteractivesimulacra}
Joon~Sung Park, Joseph~C. O'Brien, Carrie~J. Cai, Meredith~Ringel Morris, Percy Liang, and Michael~S. Bernstein.
\newblock Generative agents: Interactive simulacra of human behavior, 2023.
\newblock URL \url{https://arxiv.org/abs/2304.03442}.

\bibitem[Shinn et~al.(2023)Shinn, Cassano, Gopinath, Narasimhan, and Yao]{shinn2023reflexion}
Noah Shinn, Federico Cassano, Ashwin Gopinath, Karthik Narasimhan, and Shunyu Yao.
\newblock Reflexion: Language agents with verbal reinforcement learning.
\newblock \emph{Advances in Neural Information Processing Systems}, 36:\penalty0 8634--8652, 2023.

\bibitem[Sreedhar \& Chilton(2024)Sreedhar and Chilton]{sreedhar2024simulatinghumanstrategicbehavior}
Karthik Sreedhar and Lydia Chilton.
\newblock Simulating human strategic behavior: Comparing single and multi-agent {LLM}s, 2024.
\newblock URL \url{https://arxiv.org/abs/2402.08189}.

\bibitem[Touzel et~al.(2024)Touzel, Sarangi, Welch, Krishnakumar, Zhao, Yang, Yu, Kosak-Hine, Gibbs, Musulan, Thibault, Gurbuz, Rabbany, Godbout, and Pelrine]{touzel2024simulationsolvingsocietalscalemanipulation}
Maximilian~Puelma Touzel, Sneheel Sarangi, Austin Welch, Gayatri Krishnakumar, Dan Zhao, Zachary Yang, Hao Yu, Ethan Kosak-Hine, Tom Gibbs, Andreea Musulan, Camille Thibault, Busra~Tugce Gurbuz, Reihaneh Rabbany, Jean-François Godbout, and Kellin Pelrine.
\newblock A simulation system towards solving societal-scale manipulation, 2024.
\newblock URL \url{https://arxiv.org/abs/2410.13915}.

\bibitem[Wu et~al.(2023)Wu, Bansal, Zhang, Wu, Li, Zhu, Jiang, Zhang, Zhang, Liu, Awadallah, White, Burger, and Wang]{wu2023autogenenablingnextgenllm}
Qingyun Wu, Gagan Bansal, Jieyu Zhang, Yiran Wu, Beibin Li, Erkang Zhu, Li~Jiang, Xiaoyun Zhang, Shaokun Zhang, Jiale Liu, Ahmed~Hassan Awadallah, Ryen~W White, Doug Burger, and Chi Wang.
\newblock Autogen: Enabling next-gen {LLM} applications via multi-agent conversation, 2023.
\newblock URL \url{https://arxiv.org/abs/2308.08155}.

\bibitem[Xiao et~al.(2025)Xiao, Sun, Luo, and Wang]{xiao2025tradingagentsmultiagentsllmfinancial}
Yijia Xiao, Edward Sun, Di~Luo, and Wei Wang.
\newblock Tradingagents: Multi-agents {LLM} financial trading framework, 2025.
\newblock URL \url{https://arxiv.org/abs/2412.20138}.

\bibitem[Zhang et~al.(2024)Zhang, Lin, Sun, Qi, Yang, Chen, Lyu, Mou, Chen, Luo, Huang, Tang, and Wei]{zhang2024electionsimmassivepopulation}
Xinnong Zhang, Jiayu Lin, Libo Sun, Weihong Qi, Yihang Yang, Yue Chen, Hanjia Lyu, Xinyi Mou, Siming Chen, Jiebo Luo, Xuanjing Huang, Shiping Tang, and Zhongyu Wei.
\newblock Election{S}im: Massive population election simulation powered by large language model driven agents, 2024.
\newblock URL \url{https://arxiv.org/abs/2410.20746}.

\bibitem[Zhu et~al.(2025)Zhu, Sun, Nian, South, Pentland, and Pei]{zhu2025automatedriskygamemodeling}
Shenzhe Zhu, Jiao Sun, Yi~Nian, Tobin South, Alex Pentland, and Jiaxin Pei.
\newblock The automated but risky game: Modeling agent-to-agent negotiations and transactions in consumer markets, 2025.
\newblock URL \url{https://arxiv.org/abs/2506.00073}.

\end{thebibliography}
\bibliographystyle{colm2025_conference}

\appendix
\section{Appendix}

\subsection{Extended Related Work}
\label{appendix:extended-related-work}

This appendix briefly covers additional background work that could not be included in the main text due to space limitations.

Several recent works have leveraged large-scale multi-agent simulations to study emergent social behaviors and collective outcomes. For example, \cite{altera2024projectsidmanyagentsimulations} run large-scale simulations in Minecraft and observe emergent behavior such as specialized roles, collective rules, etc.
\cite{sreedhar2024simulatinghumanstrategicbehavior} show that multi-agent setups simulate human behavior in a game more accurately than single-agent setups. \cite{touzel2024simulationsolvingsocietalscalemanipulation} introduce a simulation system using Mastodon to study how agent interventions can manipulate opinions and election outcomes at scale. \cite{zhang2024electionsimmassivepopulation} explore an election simulation framework that involves hundreds of thousands of agents with the aim of replicating real past elections in the US.

\FloatBarrier

\subsection{Simulation Configuration Example}
\label{appendix:sim-config-example}

\begin{figure}[!tb]
\begin{lstlisting}
{
  "model": "gpt-4o",
  "config": {
    "name": "Bike Price Negotiation with Enhanced Optimization",
    "agents": [
      {
        "name": "Buyer",
        "description": "Wants the bike for the lowest possible price",
        "prompt": "You are a buyer looking to purchase a bike. Your absolute maximum is 400 Euro, but your target is as low as possible <prompt-continues ...>",
        "utility_class": "BuyerAgent",
        "strategy": {"max_price": 400},
        "self_improve": true,
        "optimization_target": true
      },
      {
        "name": "Seller",
        "description": "Selling a bike and aiming for around 400 Euro",
        "prompt": "You are selling a used bike that you think is worth around 400 Euro, but you really need to make a sale <prompt-continues ...>",
        "utility_class": "SellerAgent",
        "strategy": {"target_price": 400},
        "self_improve": false
      }
    ],
    "termination_condition": "STOP_NEGOTIATION",
    "output_variables": [
      {"name": "final_price", "type": "Number", "description": "The agreed-upon final price for the bike"},
      {"name": "deal_reached", "type": "Boolean", "description": "Whether the buyer and seller reached an agreement"},
      {"name": "negotiation_rounds", "type": "Number", "description": "Number of back-and-forth exchanges"},
      {"name": "buyer_satisfaction", "type": "Number", "description": "Buyer's satisfaction with the outcome (1-10 scale)"},
      {"name": "seller_satisfaction", "type": "Number", "description": "Seller's satisfaction with the outcome (1-10 scale)"},
      {"name": "last_offer_made", "type": "Number", "description": "The last offer made by the buyer"},
      {"name": "last_offer_received", "type": "Number", "description": "The last offer received by the seller"}
    ]
  },
  "num_runs": 10,
  "optimization_prompt": "You are an expert prompt engineer tasked with maximizing agent utility. Your goal is to rewrite the agent's prompt to achieve the HIGHEST POSSIBLE UTILITY SCORE <prompt-continues ...>",
  "simulation_context": {
    "type": "negotiation",
    "domain": "consumer_goods",
    "objectives": ["maximize_utility", "reach_agreement"],
    "constraints": ["budget_limit", "fairness"],
    "tags": ["buyer-seller", "price-negotiation", "bike-marketplace"]
  }
}
\end{lstlisting} 
\caption{Simulation configuration example\label{fig:config-example}} 
\end{figure} 

\FloatBarrier

\subsection{Reflection Prompt}
\label{appendix:reflection-prompt}

\begin{figure}
\begin{lstlisting}
prompt = (
    "You are thinking silently as " + self.name + ". "
    "In ONE short sentence, note what you believe or plan "
    "after reading:\n" + last_public_msg
)
\end{lstlisting}
\caption{Agent reflection prompt\label{fig:reflection-prompt}} 
\end{figure}

\FloatBarrier

\begin{figure}
\begin{lstlisting}
# Use custom optimization prompt if provided, otherwise use default
optimization_content = (
    "You are a seasoned negotiation coach.\n"
    f"Previous strategies:\n- "
    + "\n- ".join(environment.get(agent_strategies_key, []))
    + "\n"
    "Analyse the transcript and devise exactly ONE new negotiation strategy "
    f"sentence the {self.name} could apply in a *future* negotiation to get a better price.\n"
    "If neither party uttered 'Yes, deal!', that means no deal was reached. "
    "In that case, focus on how to reach a good deal faster next time.\n"
    "Start with an action verb and do NOT duplicate prior strategies. "
    "Do NOT mention specific prices, names or budgets from the dialogue.\n"
    f"{self._get_private_constraints()}\n."
    f"The {self.name}'s normalised utility for this deal was {utility:.2f} ({tag}).\n"
    "- If utility was 'loss' or 'poor', focus on improvement. "
    "- If 'great', suggest how to replicate or slightly enhance success. \n"
    "Include one recognised negotiation tactic (e.g., anchoring, mirroring, time-pressure) that fits what you observed in the transcript."
    "Think step-by-step and return ONLY that single negotiation strategy sentence."
)
\end{lstlisting}
\caption{Negotiation Coach Agent prompt\label{fig:negotiation-coach-prompt}} 
\end{figure}

% \subsection{Simulation Loop}

% \begin{algorithm}[h]
% \caption{Turn‑based multi‑agent simulation}
% \begin{algorithmic}[1]
% \Require Environment $E=(\text{agents},\;\text{ctx})$
% \State \textbf{repeat}
%     \ForAll{agent $A \in E.\text{agents}$ \textbf{in turn order}}
%         \State $A.\text{act}(E)$
%         \If{$A$'s stop condition holds}
%             \State \textbf{break repeat}
%         \EndIf
%     \EndFor
% \Until{termination triggered}
% \ForAll{agent $A \in E.\text{agents}$}
%     \State $s_A \gets A.\text{utility}(E, A)$ \Comment{self‑evaluation}
%     \State \textbf{output} $s_A$
% \EndFor
% \end{algorithmic}
% \end{algorithm}

\end{document}